\documentstyle[prcrc,11pt,fleqn,epsfig]{article}

\newcommand{\lesssim}{\mathrel{\rlap{\lower4pt\hbox{\hskip1pt$\sim$}}
    \raise1pt\hbox{$<$}}}         

\title{Accurate neutralino relic density\thanks{Presented by
    P. Gondolo.}}

\author{P. Gondolo\address{Max Planck Institute for Physics,
        F\"ohringer Ring 6, 80805 Munich, Germany}
        and 
        J. Edsj\"o\address{Center for Particle Astrophysics, University
          of California, \\301 Le Conte Hall, Berkeley, 
          CA 94720-7304, U.S.A.}
        }
       
\begin{document}

\null

\thispagestyle{empty}
\setcounter{page}{0}

\rightline{MPI-PhT/98-33}
\rightline{April 1998}

\begin{center}

\vspace{2cm}

{\LARGE\bf Accurate neutralino relic density}

\vspace{1cm}

{\large P. Gondolo$^{\rm a}$ and J. Edsj\"o$^{\rm b}$ }

\vspace{0.5cm}

${}^{\rm a}$ {\it Max Planck Institute for Physics,
        F\"ohringer Ring 6, 80805 Munich, Germany}

${}^{\rm b}$ {\it Center for Particle Astrophysics, University
          of California, \\301 Le Conte Hall, Berkeley, 
          CA 94720-7304, U.S.A.}

\vspace{1cm}

\end{center}

\begin{abstract}
  We enlarge our set of supersymmetric models and update accelerator
  constraints in our precise calculation of the relic density of the lightest
  neutralino, which includes relativistic Boltzmann averaging,
  subthreshold and resonant annihilations, and coannihilation processes among
  charginos and neutralinos.
\end{abstract}

\vspace{\fill}

\begin{center}

{\large\it Presented by P. Gondolo at Dark Matter 98, Marina del Rey,
  California, February 1998. }

\vspace{2cm}
       
\end{center}

\newpage

\maketitle

\begin{abstract}
  We enlarge our set of supersymmetric models and update accelerator
  constraints in our precise calculation of the relic density of the lightest
  neutralino, which includes relativistic Boltzmann averaging,
  subthreshold and resonant annihilations, and coannihilation processes among
  charginos and neutralinos.
\end{abstract}

\section{Introduction}

The lightest neutralino is one of the most promising candidates for the dark
matter in the Universe. A linear combination of the superpartners of the
neutral gauge and Higgs bosons, it is believed to be the lightest stable
supersymmetric particle in the Minimal Supersymmetric extension of the Standard
Model (MSSM). 

In the near future, high precision measurements of the dark matter density may
become possible from high resolution maps of the cosmic microwave background,
and this may lead to constraints on supersymmetry. It is therefore of great
interest to calculate the relic density of the lightest neutralino as
accurately as possible.

As a major step towards a complete and precise calculation valid for all
neutralino masses and compositions, we include
all concomitant annihilations (coannihilations) 
between neutralinos and charginos, properly
treating thermal averaging in presence of thresholds and resonances in the
annihilation cross sections (for details see \cite{EG}).


\section{Formalism}
\label{sec:Boltzmann}

Consider coannihilation of $N$ supersymmetric particles with masses $m_i$ and
statistical weights $g_i$ (first studied in ref.~\cite{Salati}).  Normally,
all heavy particles have time to decay into the lightest one, which we assume
stable. Its final abundance is then simply described by the sum of the
densities $ n= \sum n_{i}.$ When the scattering rate of supersymmetric
particles off the thermal background is much faster than their annihilation
rate, $n$ obeys the evolution equation \cite{GriestSeckel}
$
  {dn}/{dt} =
  -3Hn - \langle \sigma_{\rm{eff}} v \rangle 
  ( n^2 - n_{\rm{eq}}^2 )
$
with effective annihilation cross section
$
  \langle \sigma_{\rm{eff}} v \rangle = 
  A/n_{\rm{eq}}^2 $.
The numerator $A$ is the total annihilation rate per unit volume at temperature
$T$, and $n_{\rm eq}$ is the total equilibrium density. We find~\cite{EG}
$
  A = (T/16 \pi^4) \int_{4m_\chi^2}^\infty \!ds
  \sqrt{s-4m_\chi^2} \, W(s) \,
  K_{1}\!(\sqrt{s}/T) 
$
and
$
  n_{\rm eq} = 
  (T/2\pi^2) \sum_i g_i m_{i}^2
  K_{2}\!( m_{i}/T ) .
$
Here $K_i(x)$ is the modified Bessel function of the second kind of 
order $i$, and
$
  W(s) = \sum_{ij}  
  g_i g_j \sigma_{ij} 
  \, \lambda(s,m_i^2,m_j^2) / 2\sqrt{\lambda(s,m_\chi^2,m_\chi^2)}
$
with $\lambda(x,y,z) = x^2+y^2+z^2-2xy-2xz-2yz$.
$W(s)$ is a Lorentz invariant annihilation rate per unit volume in which
coannihilations appear as thresholds at $\sqrt{s}$ equal to the sum of the
masses of the coannihilating particles.  The independence of $W(s)$ on
temperature is a remarkable calculational advantage in presence of
coannihilations: in fact it can be tabulated in advance, before taking the
thermal average and solving the density evolution equation.

The previous equations generalize the result of
Gondolo and Gelmini~\cite{GondoloGelmini} to coannihilations.


\section{Results}
\label{sec:Results}

To explore a significant fraction of the MSSM parameter space \cite{haberkane},
we keep the number of theoretical relations among the parameters to a minimum.
We assume GUT relations for gaugino masses, keep only the top and bottom
trilinear soft supersymmetry-breaking parameters, and use a single mass
parameter for the diagonal entries in the sfermion mass matrices at the weak
scale. We perform many different scans in parameter space, some general and
some specialized to interesting regions.  We keep only models that satisfy the
experimental constraints on the $Z^0$ width, on the $b \rightarrow s \gamma$
branching ratio, and on superpartner and Higgs boson masses. Here we show
results for $\sim$85,000 supersymmetric models that satisfy accelerator
constraints as of March 1998 (we include the ALEPH bounds at $\sqrt{s}
= 184$~GeV~\cite{JCarr}).

\begin{figure}[htb]
\begin{minipage}[t]{77.5mm}
  \vspace{-0.2in}
  \epsfig{file=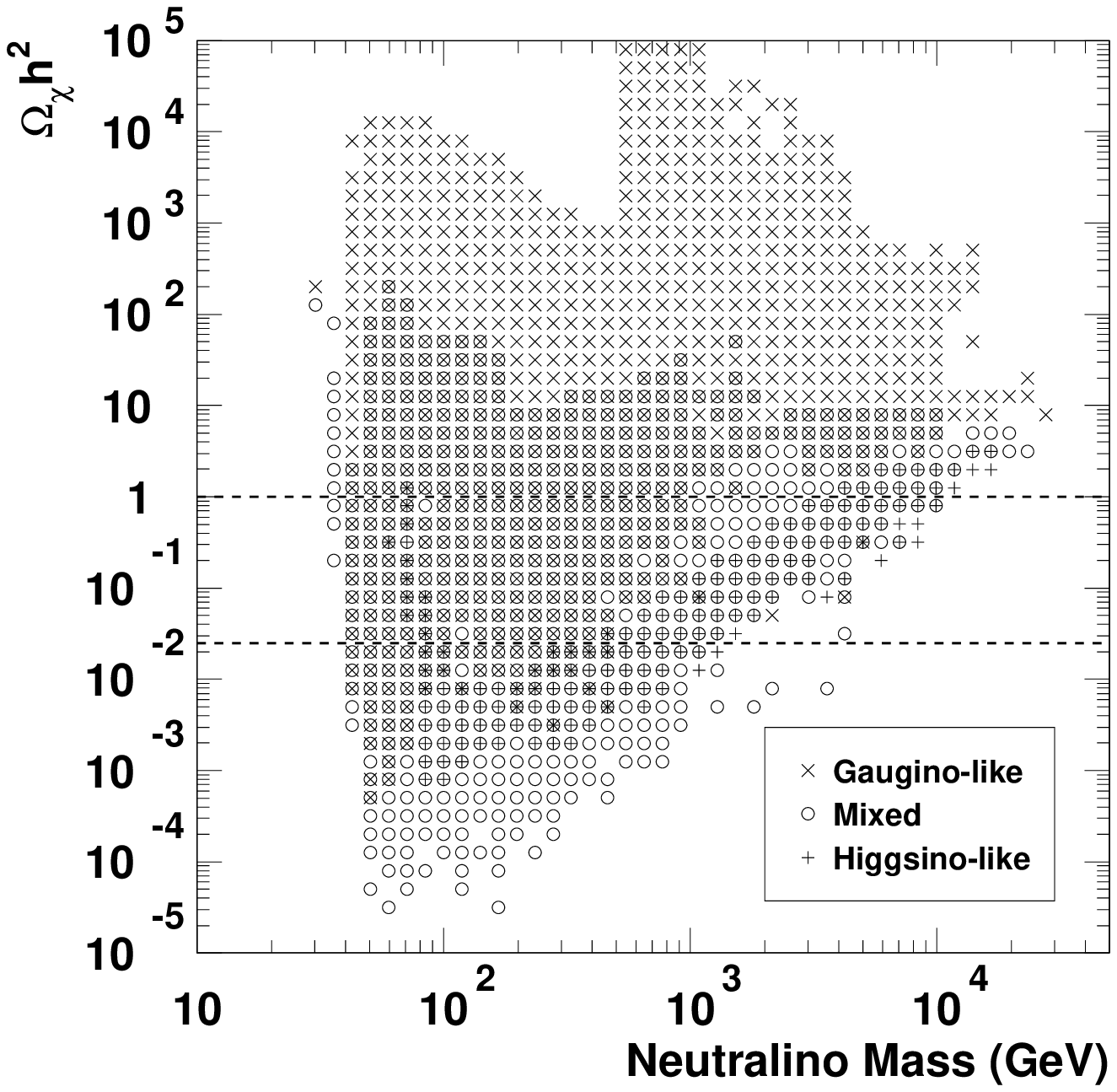,width=85mm}
  \vspace{-0.5in}
  \caption{
    Neutralino relic density versus neutralino mass.  Horizontal lines
    show the cosmologically interesting region $0.025 < \Omega_\chi
    h^2 <1$.}
  \label{fig:oh2vsmx}
\end{minipage}
\hspace{\fill}
\begin{minipage}[t]{77.5mm}
  \vspace{-0.2in}
  \epsfig{file=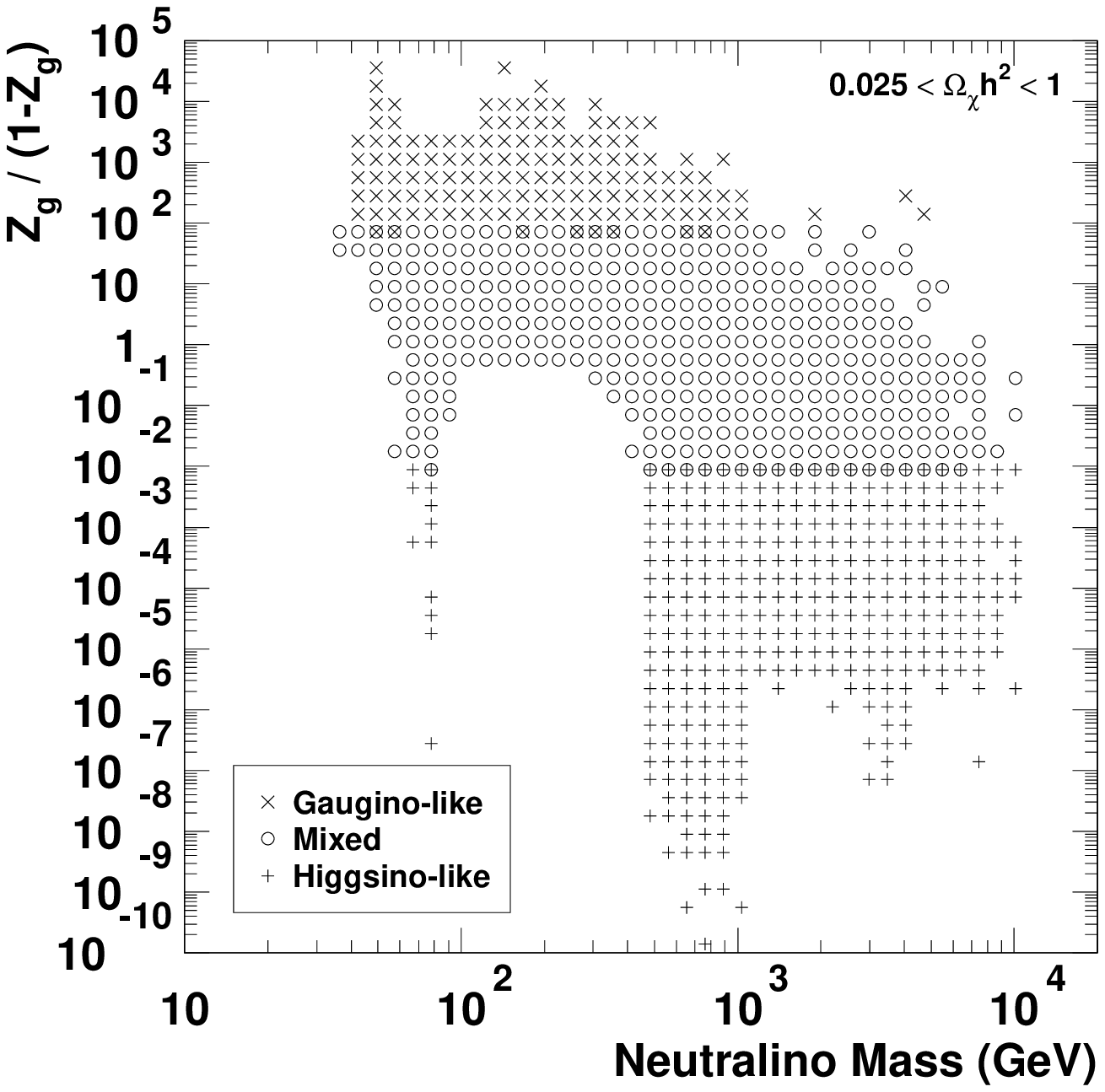,width=85mm}
  \vspace{-0.5in}
  \caption{Neutralino masses $m_\chi$ and compositions $Z_g/(1-Z_g)$
    for cosmologically interesting models.}
  \label{fig:cosmregion}
\end{minipage}
\end{figure}

We obtain analytic expressions for the many Feynman diagrams contributing to
the two-body cross sections at tree level for
neutralino-neutral\-ino, neutralino-chargino and chargino-chargino
annihilation. Then for each set of model parameters, we sum over particle
polarizations and over initial and final states numerically and tabulate the
annihilation rate $W(s)$. Thermal averaging and integration of the density
equation finally give the neutralino relic density $\Omega_\chi h^2 = m_\chi
n_0/\rho_{\rm crit}$ in units of the critical density $\rho_{\rm crit}$.

We find that coannihilation processes are important not only for light
higgsino-like neutralinos, as pointed out
before in approximate calculations~\cite{MizutaYamaguchi}, but also for heavy
higgsinos and for mixed and gaugino-like neutralinos.  Indeed, coannihilations
should be included whenever $|\mu| \lesssim 2 |M_1|$, independently of the
neutralino composition. When $|\mu| \sim |M_1|$, coannihilations can increase
or decrease the relic density in and out of the cosmologically interesting
region.

Fig.~\ref{fig:oh2vsmx} shows the neutralino relic density $\Omega_\chi h^2$
versus the neutralino mass $m_\chi$. To avoid artificial bands and holes in the
point distribution, we have divided the plotted region into cells and marked
those in which we find at least one model allowed by present accelerator
constraints.
The horizontal lines limit the cosmologically interesting region where the
neutralino can constitute most of the dark matter in galaxies without violating
the constraint on the age of the Universe. We take it to be $0.025 <
\Omega_{\chi} h^2 < 1$.

Fig.~\ref{fig:cosmregion} shows the cosmologically interesting region in the
neutralino mass--composition plane ($Z_g$ is the gaugino fraction). 
This region is limited to the left by accelerator
constraints, to the right by $\Omega_\chi h^2 < 1$, and below and above by
incompleteness in our survey of parameter space, except for the hole at 85--450
GeV where $\Omega_\chi h^2 < 0.025$.

The main effect of coannihilations is to shift the higgsino region to higher
masses. In particular the cosmological upper bound on the neutralino mass
becomes 10 TeV.  Differently from previous approximate
results~\cite{MizutaYamaguchi}, we find a cosmologically interesting
window of light higgsino-like neutralinos with masses around 75 GeV.

We conclude that if coannihilations are properly included, the neutralino is a
good dark matter candidate whether it is light or heavy, and whether it is
higgsino-like, mixed or gaugino-like.


\end{document}